\documentclass[review]{elsarticle}

\usepackage{hyperref}

\journal{Renewable Energy}

\DeclareGraphicsExtensions{.ps, .png, .eps, .tiff}

\makeatletter
\newcommand{\thickhline}{%
    \noalign {\ifnum 0=`}\fi \hrule height 0.75pt
    \futurelet \reserved@a \@xhline
}
\makeatother

\makeatletter
\newcommand*{\rom}[1]{\expandafter\@slowromancap\romannumeral #1@}
\makeatother
\usepackage{supertabular}
\usepackage{array}
\usepackage{setspace}
\newenvironment{myenv}[1]
{\begin{spacing}{#1}}
    {\end{spacing}}
\usepackage{anysize}
\usepackage{amsmath}
\usepackage{graphicx}
\usepackage{eurosym}
\usepackage{microtype}
\usepackage{amssymb}
\usepackage{yfonts}
\usepackage{url}
\usepackage{multirow}
\usepackage{xcolor}

\marginsize{20mm}{20mm}{20mm}{25mm}

\begin{document}
\begin{frontmatter}

\title{Technical Barriers for Harnessing the Green Hydrogen: A Power System Perspective}

\author{Abbas Rabiee\corref{mycorrespondingauthor}
\address{School of Electrical and Electronic Engineering, University College Dublin, Ireland}}

\cortext[mycorrespondingauthor]{Corresponding author}
\ead{abbas.rabiee@ucd.ie}

\author{Andrew Keane
\address{School of Electrical and Electronic Engineering, University College Dublin, Ireland}}

\author{Alireza Soroudi
\address{School of Electrical and Electronic Engineering, University College Dublin, Ireland}}

\begin{abstract}
Extracting green hydrogen from renewable energy sources is a new concept in the energy industry. As an energy carrier, hydrogen is well capable of facilitating a strong coupling between various energy sectors, as well as integration of renewable energy sources. This paper investigates the system-wide technical factors that might limit the amount of producible hydrogen in a given power system. 
A non-linear programming formulation is proposed to quantify the impact of voltage security constraints, the location and size of power to hydrogen facilities, and finally the wind penetration levels on the harvest-able green hydrogen. The applicability of the proposed framework is demonstrated on the IEEE 39 bus system. 
\end{abstract}

\end{frontmatter}

\section{NOMENCLATURE}
The notations and symbols used throughout the paper are stated in this section.
\begin{myenv}{1.1}
    \begin{supertabular}{>{\arraybackslash}p{1.8cm} >{\arraybackslash}p{13cm} }
        \textbf{Abbreviations:} & \\
        $\textit{OPF}$&\textcolor{black}{Optimal power flow.}\\  
        $\textit{COP}$&Current operation point.\\
        $\textit{SLP}$&Security limit point.\\
        $\textit{LM}$&Loading margin.\\
        $\textit{P2H}$& Power to hydrogen.\\
        $\textit{WF}$& Wind farm.\\
        $\textit{RES}$&Renewable energy source.\\
        \vspace{1mm}
        \textbf{Sets/Indices:} & \\
        $c$&Index for operation point ($c_0$ denotes COP, $c_1$ denotes SLP).\\        
        $b,k$&Index for electrical network buses.\\
        $t$&Index for operation intervals.\\
        $\Omega_B$ & Set of network all buses.\\
        $\Omega_{B_G}$&Set of network buses with generators.\\
        $\Omega_{B_b}$&Set of network buses connected to bus $b$.\\
        $\Omega_{B_W}$&Set of network buses connected to WFs.\\
        $\Omega_{B_{P2H}}$&Set of network buses connected to P2H units.\\
        $\Omega_{T}$&Set of time intervals.\\
        \textbf{Variables:}&\\
        $TH$&  Total hydrogen extracted in the entire interval, $(kg)$.\\
        $H_{b,t}$& Hydrogen extracted from the electrolyser connected to bus $b$ at time $t$, $(kg/h)$.\\
        $(PW/QW)_{b,t}$&Active/reactive power of WF in bus $b$ at time $t$, (pu).\\
        $(PG/QG)_{b,t,c}$&Active/reactive power generation in node $b$ at time $t$ and operation point $c$, (pu).\\
        $(PD/QD)_{b,t,c}$&Active/reactive demand in bus $b$ at time  $t$ and operation point $c$, (pu).\\    
        $(PH/QH)_{b,t,c}$&Active/reactive power demand of P2H unit in bus $b$ at time $t$, (pu).\\
        $QG^{L_1/L_2}_{b,t,c}$& Reactive power limit corresponding to stator/field current limit in bus $b$, time $t$ and operation point $c$, (pu)\\
        $(\overline{QG}/\underline{QG})_{b,t,c}$& Upper/lower limit of reactive power generation in bus $b$, time $t$ and operation point $c$, (pu).\\
        $(y/z)_{b,t}$&Auxiliary variables for modeling active and reactive power limits of generators.\\
        $v^{up/dn}_{b,t}$&Auxiliary variables for modelling the voltage difference of COP and SLP as a result of reactive power limit activation.\\
        $\lambda$& Loading parameter. \\
        $S_{bk,t,c}$& Apparent power flowing through $bk$-th line at time $t$ and operation point $c$, (pu).\\
        $(V/\theta)_{b,t,c}$&Voltage magnitude/angle of bus $b$ at time $t$ and operation point $c$, (pu/rad).\\
        \textbf{Parameters:}&\\
        $K^{P/Q}_{b}$ &Active/reactive power demand increment factor at bus $b$.\\
        $K^{G}_{b}$ &Active power generation increment factor at bus $b$.\\
        $Y_{bk}/\gamma_{bk}$&Magnitude/angle of $bk$-th element of system admittance matrix.\\
        $\eta_{b,t}$&Efficiency factor of the electrolyser supplied from bus $b$ at time $t$, $kg/MWh$.\\
        $E_{b,t,c}$&Internal voltage of the generator connected to bus $b$, time $t$ and operation point $c$, (pu).\\
        $X_{s_b}$&Synchronous reactance of the generator connected to bus $b$.\\
        $\overline{\delta}_{b}$&Upper limit on angular separation of stator and rotor fields of the generator connected to bus $b$, (rad).\\
        $\overline{IG}_{b}$&Upper limit on the stator current of the generator connected to bus $b$, (pu)  \\
        $M_{1/2,b}$& Big-M constants for reactive/active power generation limits modelling at bus $b$.\\
        ${\overline{X}/\underline{X}}$& Upper/lower bound of variable $X$\\
    \end{supertabular}
\end{myenv}

\vspace{2mm}
\section{Introduction}\label{sec1}
\subsection{Background and Motivations}
\indent Hydrogen produced through renewable energy sources (RESs), known as green hydrogen, can provide clean energy to the main economy sectors such as industry, buildings, and transport \cite{staffell2019role,wang2020well}. In this way, the goal of $40\%$ share of electricity as the dominant energy carrier in 2050 would be realized and hence, the decarbonised energy world envisaged by the Paris Agreement will likely be reachable \cite{IRENA1}. \textcolor{black}{An example of green hydrogen impact on carbon intensity reduction in industry sector is reported by \cite{jiao2020green}.}
Power systems, as the main infrastructure of large-scale electricity generation and transmission, are forced to be operated even closer to their security limits due to the increasing demand, market pressures, public and decarbonization concerns.

\begin{figure}[ht]
  \centering
  \includegraphics[width=0.5\columnwidth,bb=60 40 520 800,angle=90, clip=true]{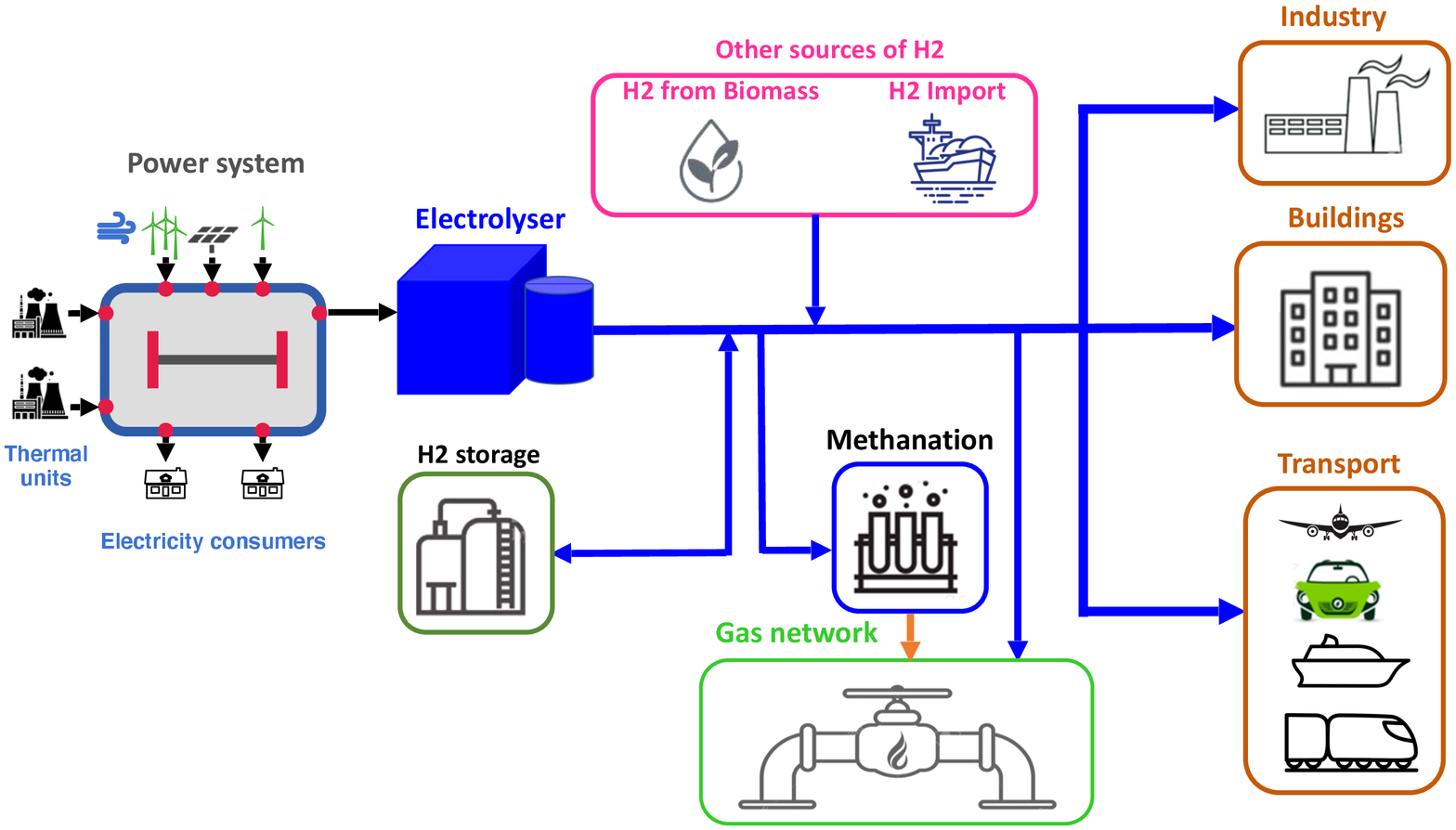} 
  \vspace{1mm}
      \caption{The link between various energy sectors via green hydrogen.}
      \label{fig:H2}
\end{figure}

Green hydrogen energy carrier facilitates large amounts
of renewable energy to be directed from the
power systems into the end-use sectors such as transport, buildings and industry, as shown in Fig. \ref{fig:H2}. Green hydrogen is extracted from water through an electrolyser consuming electric energy \cite{FASIHI2020118466}. Hence, integration of RESs and large-scale energy storage in mid and long-term intervals could be accomplished via green hydrogen. 
\textcolor{black}{The future carbon-less energy chain via green hydrogen can be obtained by the following means:}
\begin{itemize}
    \item \textcolor{black}{By production of green hydrogen, the need for blue hydrogen (i.e. the hydrogen extracted from natural gas through the process of steam methane reforming) and hence the CO2 generation will be reduced.}
    \item \textcolor{black}{The generated green hydrogen could be injected to natural gas network (as shown in Fig. \ref{fig:H2}) up to a specific percentage. By this ability, the amount of natural gas consumption will be decreased with respect to the case of no green hydrogen.}
    \item \textcolor{black}{Green hydrogen could be stored and used in marine, aviation and other transportation systems through the hydrogen supply chain.}
\end{itemize}

Electrolysers technology has been matured in past decades and their scale-up in near future can increase the share of green hydrogen in the decarbonised energy chain, worldwide. Cost-benefit analyses done on this technology reveal that it will be a competitive energy carrier in the near future \cite{eichman2016economic, hou2017optimizing, kopp2017energiepark}. Electrolysers, as the integral part of green hydrogen technology, offer a flexible load to the power systems that can easily provide ancillary services such as grid balancing services (upwards and downwards frequency regulation) at the same time operating at optimal capacity to meet demand for hydrogen from other downstream hydrogen energy sectors (such as industry, buildings and transport). Hence, introducing the green hydrogen (as a new flexible load) to the future power systems has different positive and negative aspects. 
On the positive side, reduction in wind curtailment (stored as hydrogen), enhancing the decarbonisation strategies can be observed. However, the level of stress will be higher since the green hydrogen will be treated as a new electric load. This new load will further limit the available power transfer capacity (ATC) of transmission networks.\\
\indent The ATC is constrained by voltage security to avoid the voltage collapse in heavy loading circumstances or contingencies.
It is desirable to operate the system with a sufficient loading margin (LM) such that the system can survive the collapse under heavy loading conditions as well as any significant single or multi-component outages.
Green hydrogen, i.e. the hydrogen that is produced from RESs could be extracted via an electrolysis process that takes place in electrolyzers. This power to hydrogen (P2H) energy conversion path could facilitate the integration of high levels of RESs into the emerging energy system. It can also provide grid balancing
services and long-term storage to manage the variation in power supply from intermittent RESs \cite{mansour2020robust, MURRAY2020117792, IRENA2}.

\subsection{Contributions}
\textcolor{black}{Large-scale green hydrogen extraction from emerging power systems with high penetration of RESs, can pose various impacts on the planning and operation of such networks, which can be considered as a research gap. As an example, incurring a large amount of electricity via P2H electrolyzers can push the system closer to its security boundaries. Hence, in this paper, the main factors affecting the harvestable hydrogen from electric power systems with high penetration of RESs, are discussed. Particularly, the focus of this research will be on investigating the impact of the following factors on the exploitable green hydrogen:
\begin{itemize}
    \item Allocation of large-scale P2H facilities (including their size and location) in the transmission system.
    \item Considering the steady-state voltage security constraints.
    \item Considering high penetration levels of RESs. 
\end{itemize}  
}

\subsection{Paper organization}
The remainder of this paper is organized as follows: Section \ref{PM} describes the proposed model for hydrogen harnessing from RESs by considering the voltage security constraints. Section \ref{sec:simulation} shows the simulation results. Finally, the paper concludes in Section \ref{sec:Conclusion}.

\section{Problem statement}
\label{PM}
\subsection{Conceptual interpretation}
\label{bcpm}
\begin{figure}[th]
  \centering
  \includegraphics[width=0.55\columnwidth, clip=true]{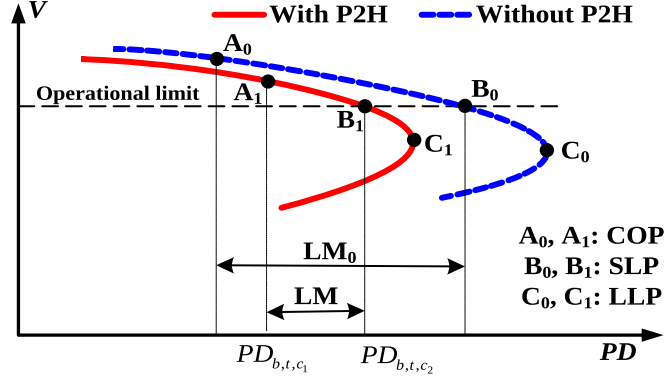} 
  \vspace{1mm}
      \caption{P-V curve to characterize the impact of P2H demand.}
      \label{fig:cop}
\end{figure}

\textcolor{black}{Increasing the system demand beyond its ATC, as well as the inability of the system to meet reactive demand are recognized as the main factors jeopardizing the voltage security of power systems \cite{6294475}. In order to characterize the loadability, as well as the voltage security, the concept of LM is employed. LM is the amount of load increase not arousing the voltage collapse or violation of critical operational constraints (such as voltage, reactive power and line flow limits). By preserving a proper level of LM, i.e. the distance between the current operation point (COP) and the security limit point (SLP), the voltage security of system will be ensured \cite{8606272}. As it is aforementioned, integration of large-scale P2H demand to generate green hydrogen, can be interpreted as the stressing the system toward its SLP. This issue is conceptually demonstrated in Fig. \ref{fig:cop}, where the loci of steady state operation points of the system with/without P2H demands are compared. It can be observed that the COP and SLP are initially $A_0$ and $B_0$, with a LM equals to $LM_0$. Integration of P2H demand changes the COP and SLP to $A_1$ and $B_1$, respectively, with the corresponding security margin of $LM_1$.}
 \vspace{2mm}

\subsection{Formulation of green hydrogen harvesting problem}
\label{sec:Information}
\textcolor{black}{The aim is to maximize the total extracted hydrogen (TH) via P2H units at the entire operating horizon (e.g. 24 hours), as follows.
\setlength{\arraycolsep}{0.0em}
\begin{alignat}{2}
\label{TCt}
& \max_{DV} \ \  TH=\sum_{t}\sum_{b \in \Omega_{B_{P2H}}}{ H_{b,t} } \\
& Subject \textit { to : } \eqref{p2h} \textit { to  }\eqref{n38} \notag
\end{alignat}
The efficiency of electrolyser depends on its loading level, such that in lower loading levels, the efficiency is higher. Thus, the extractable hydrogen (in $kgH_2$ per $MWh$ of input electricity) is a function of electrolyser's loading level. This functionality can be expressed as follows at the COP of the system.
\begin{alignat}{2}
\label{p2h}
& H_{b,t}=f\left(PH_{b,t,c_1} \right) \ \ \forall b \in \Omega_{B_{P2H}} 
\end{alignat}
where $f(.)$ is a nonlinear function \cite{IRENA1}. The existing electrolyser technologies, i.e. alkaline and proton exchange membrane, have almost a constant efficiency when operating near their nominal capacity \cite{IRENA1}. Since it is assumed that the electrolysers are connected to the network and sufficient downstream hydrogen demand as well as hydrogen storage capacity exist, assuming that the electrolysers are operated near their nominal values is rational. Hence, one can consider a constant conversion ratio between its output hydrogen and input electric energy as follows.
\begin{alignat}{2}
\label{p2h2}
& H_{b,t}=\eta_{b,t} PH_{b,t,c_1}  \ \ \forall b \in \Omega_{B_{P2H}} 
\end{alignat}
where $\eta_{b,t}$ in \eqref{p2h2} is the efficiency factor of the electrolyser (in $kg/MWh$), supplied from bus $b$ at time $t$.
}

\setlength{\arraycolsep}{5pt}
\textcolor{black}{The AC power flow equations are considered for COP and SLP, simultaneously. Also, since the reactive power support by generators plays a crucial role in voltage security, in this paper, a detailed model of capability curve of generators is considered by including the field current, armature current and under excitation limits. Besides, without loss generality, the wind power is considered as the available RES at the system.}


\begin{alignat}{2}
\label{lf1}
& PG_{b,t,c}+PW_{b,t,c}-PD_{b,t,c}-PH_{b,t,c}= \sum_{k \in \Omega_{B_b}} V_{b,t,c}V_{k,t,c} Y_{bk}cos\left(\theta_{b,t,c} -\theta_{k,t,c} - \gamma_{bk}\right)  \forall b \in \Omega_B  \\
\label{lf2}
& QG_{b,t,c}-QD_{b,t,c}-QW_{b,t,c}-QH_{b,t,c}=  \sum_{k \in \Omega_{B_b}} V_{b,t,c}V_{k,t,c}  Y_{bk}sin\left(\theta_{b,t,c} -\theta_{k,t,c} - \gamma_{bk}\right) \forall b \in \Omega_B 
\end{alignat}
where \eqref{lf1} and \eqref{lf2} are hourly nodal active and reactive power balances in COP and SLP. Active and reactive power demands of electrolyser, i.e. $P2H_{b,t,c}$ and $Q2H_{b,t,c}$, are included in these equations. 

In order to preserve the adequacy of network, the reserve constraint is considered as follows.
\begin{alignat}{2}
\label{lf3}
& \sum_{k_{\neq b} \in \Omega_{B_G}}  (\overline{PG}_{k}-PG_{k,t,c_1}) \ge PG_{b,t,c_1} \ \ \ \ \ \ \ \ \forall b \in \Omega_{B_G} 
\end{alignat}
where \eqref{lf3} is the hourly reserve constraint, that ensures an available generation capacity greater than power output of the largest running generating unit.

Despite of various positive impacts of RESs (such as wind energy), their penetration level is usually limited due to stability issues. This limit is considered as follows, as a percentage of the hourly demand.
\begin{alignat}{2}
\label{lf4}
& \sum_{b \in \Omega_{B_W}} PW_{b,t,c_1} \le \alpha \times \sum_{b \in \Omega_{B}} PD_{b,t,c_1}  
\end{alignat}
The hourly wind power penetration in \eqref{lf4} can be regulated by $0 \le \alpha \le 1$. 

Also, since a multi-period optimal power flow model is developed in this paper, the ramp-rate constraints of the generators should be considered as follows.
\begin{alignat}{2}
\label{lf5}
& -RD_{b} \le \left(PG_{b,t,c}-PG_{b,t-1,c} \right) \le RU_{b} \ \ \ \ \  \forall b \in \Omega_{B_G}
\end{alignat}
Voltage security of the network is highly depends on the reactive power generation capability of generators. Hence, for the sake of accurate characterisation of voltage security, accurate model of generators capability curve is necessary. In this paper the capability curves are modeled via \eqref{lf6A}-\eqref{lf6D}, which express respectively the armature current, field current and under excitation limits, respectively. It is worth to note that the maximum reactive power supply by each generator for a given active power generation ($PG_{b,t,c}$), is determined from \eqref{lf6D}.
\begin{alignat}{2}
\label{lf6A}
&\left(PG_{b,t,c}\right)^2+\left( QG^{L_1}_{b,t,c} \right)^2=\left(V_{b,t,c} \overline{IG}_{b}\right)^2\\
\label{lf6B}
&\left(PG_{b,t,c}\right)^2+\left( QG^{L_2}_{b,t,c}+\frac{V^{2}_{b,t,c}}{X_{s_b}} \right)^2=\left( \frac{V_{b,t,c}E_{b,t,c}}{X_{s_b}} \right)^2\\
\label{lf6C}
& \underline{QG}_{b,t,c}=PG_{b,t,c} cot(\overline{\delta}_{b})-\frac{V^{2}_{b,t,c}}{X_{s_b}}\\
\label{lf6D}
&\overline{QG}_{b,t,c}=\min\left(QG^{L_1}_{b,t,c}, QG^{L_2}_{b,t,c}\right)
\end{alignat}

Also, \eqref{lf7}-\eqref{lf12} are the upper/lower limits of active/reactive power generations, voltages, WFs' active/reactive power injections, P2H active/reactive power demand and apparent power flowing through lines, respectively.

\begin{alignat}{2}
\label{lf7}
& \underline{PG}_{b} \le PG_{b,t,c} \le  \overline{PG}_{b} \hspace{3.2cm}  \forall b \in \Omega_{B_G}\\
\label{lf8}
& \underline{QG}_{b,t,c} \le QG_{b,t,c} \le  \overline{QG}_{b,t,c} \hspace{2.4cm} \forall b \in \Omega_{B_G}\\
\label{lf9}
& \underline{V}_{b} \le V_{b,t,c} \le  \overline{V}_{b} \hspace{4cm} \forall b \in \Omega_{B}\\
\label{lf10}
& 0 \le PW_{b,t,c} \le  \overline{PW}_{b,t} \hspace{3.3cm} \forall b \in \Omega_{B_W}\\
\label{lf11}
& \underline{QW}_{b,t} \le QW_{b,t,c} \le  \overline{QW}_{b,t} \hspace{2.5cm} \forall b \in \Omega_{B_W}\\
\label{lf11_1}
& \underline{PH}_{b} \le PH_{b,t,c} \le \overline{PH}_{b} \hspace{2.8cm} \forall b \in \Omega_{B_{P2H}}\\
\label{lf11_2}
& \underline{QH}_{b} \le QH_{b,t,c} \le \overline{QH}_{b} \hspace{2.8cm} \forall b \in \Omega_{B_{P2H}}\\
\label{lf12}
& \underline{S}_{bk} \le S_{bk,t,c} \le  \overline{S}_{bk} \hspace{2.3cm} \forall b \in \Omega_{B}, \forall k \in \Omega_{B_b}
\end{alignat}

As it is aforementioned, in order to characterise voltage security issue, loading parameter (i.e. $\lambda$ ) is used to scale the demands and generations from COP to SLP (denoted by $c_1$ and $c_2$, respectively). The relationship between COP and SLP is expressed by \eqref{n28}-\eqref{n30}.
\begin{alignat}{2}
\label{n28}
&PD_{b,t,c_2}=
\begin{pmatrix}
1+K^{P}_{b}\lambda\,
\end{pmatrix}
\times PD_{b,t,c_1} \hspace{1.8cm} \forall b \in \Omega_{B}\\
\label{n29}
&QD_{b,t,c_2}=
\begin{pmatrix}
1+K^{Q}_{b}\lambda\,
\end{pmatrix}
\times QD_{b,t,c_1}\hspace{1.8cm} \forall b \in \Omega_{B}\\
\label{n30}
&PG_{b,t,c_2}=\min((1+K^{G}_{b}\lambda )\times PG_{b,t,c_1},\overline{PG}_{b}) \hspace{0.05cm} \forall b \in \Omega_{B_G}
\end{alignat}

Also, (\ref{n35})-(\ref{n38}) indicate that for generators that reach their reactive power limits, voltage magnitude at the corresponding bus is not equal in COP and SLP. It is worth to note that $\overline{QG}_{b,t,c_2}$ and $\underline{QG}_{b,t,c_2}$ in \eqref{n36} and \eqref{n37} are defined in \eqref{lf6C} and \eqref{lf6D}, respectively.
\begin{alignat}{2}
\label{n35}
&V_{b,t,c_2}=V_{b,t,c_1}+ v^{dn}_{b,t}-v^{up}_{b,t}\\
\label{n36}
&(\overline{QG}_{b,t,c_2}-QG_{b,t,c_2}) \times v^{up}_{b,t} \le 0 \\
\label{n37}
&(QG_{b,t,c_2}-\underline{QG}_{b,t,c_2}) \times v^{dn}_{b,t} \le 0 \\
\label{n38}
&v^{dn}_{b,t}, v^{up}_{b,t} \geq 0 
\end{alignat}
\subsection{Reformulation of complicated constraints}
\label{ssec:reformulat}
It is evident from \eqref{lf6D} and \eqref{n30} that the $\min$ operator is a non-differentiable function resulting a mixed integer NLP (MINLP) optimization model. In order to avoid dealing with such an MINLP model, the equation \eqref{lf6D} is replaced by \eqref{n31A1}-\eqref{n31E1}, and \eqref{n30} is substituted by \eqref{n31A}-\eqref{n31E}.
\begin{alignat}{2}
\label{n31A1}
& QG_{b,t,c_2} \le  QG^{L_1}_{b,t,c_1} \\ \label{n31B1}
& QG_{b,t,c_2} \le QG^{L_2}_{b,t,c_1} \\
\label{n31C1}
& QG_{b,t,c_2} \ge QG^{L_1}_{b,t,c_1} - M_{1,b} \times y_{b,t} \\
\label{n31D1}
& QG_{b,t,c_2} \ge QG^{L_2}_{b,t,c_1}-  M_{1,b} \times (1-y_{b,t})\\
\label{n31E10}
& y_{b,t} = \left(y_{b,t}\right)^{2}\\
\label{n31E1}
& 0 \le y_{b,t} \le 1\\
\label{n31A}
& PG_{b,t,c_2} \le (1+K^{G}_{b}\,\lambda ) \times P^{G}_{b,t,c_1} \\ 
\label{n31B}
& PG_{b,t,c_2} \le \overline{PG}_{b} \\
\label{n31C}
& PG_{b,t,c_2} \ge (1+K^{G}_{b}\,\lambda ) \times PG_{b,t,c_1} - M_{2,b} \times z_{b,t} \\
\label{n31D}
& PG_{b,t,c_2} \ge \overline{PG}_{b}- M_{2,b} \times (1-z_{b,t})\\
\label{n31E11}
& z_{b,t} = \left(z_{b,t}\right)^{2}\\
\label{n31E}
& 0 \le z_{b,t} \le 1
\end{alignat}

\subsection{Decision variables}
\textcolor{black}{The sets of decision variables ($DV$) of the proposed optimisation model is as follows.}
\setlength{\arraycolsep}{0.0em}
\begin{alignat}{2}
\label{n44}
DV=&\begin{Bmatrix}
    V_{b,t,c}, (PG/QG)_{b,t,c}, (PW/QW)_{b,t,c}\\
    P2H_{b,t,c}, QG^{L_1/L_2}_{b,t,c},
    \lambda, y_{b,t}, z_{b,t}
    \end{Bmatrix}
\end{alignat}



\section{Numerical Studies}
\label{sec:simulation}
\subsection{Data}
\begin{figure}[ht]
  \centering
  \includegraphics[width=0.8\columnwidth,bb=0 0 950 630,angle=0, clip=true]{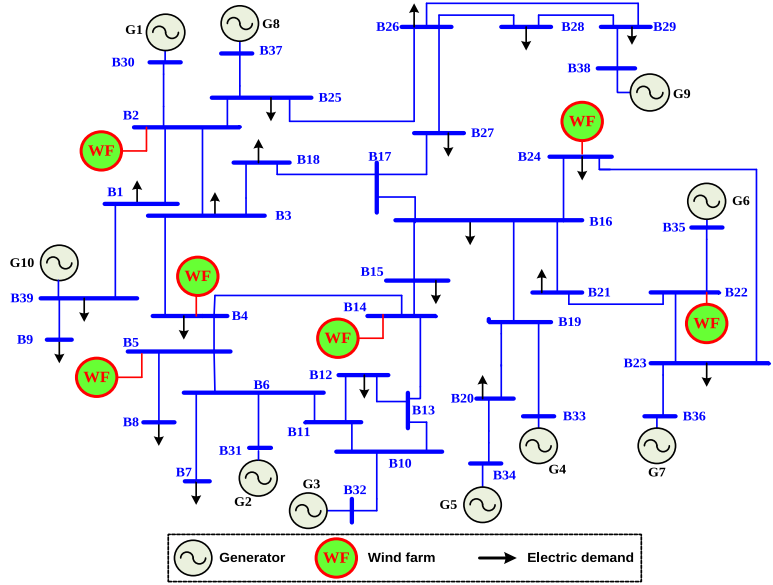} 
  \caption{\textcolor{black}{Single-line diagram of the studied network.}}
      \label{fig:slgnetwork}
\end{figure}

In order to study the influencing factors on large-scale green hydrogen extraction via power systems, the aforementioned model is implemented on the IEEE 39-bus standard system. The one-line diagram of this network is depicted in Fig. \ref{fig:slgnetwork}. Also, the network data is given in \cite{zimmerman2010matpower}. The internal voltage and synchronous reactance of all generators are assumed to be $2.574 pu$ and $1.912 pu$ on each generator's rated power basis, respectively. Besides, $M_{1/2,b}$ are assumed to be twice the $\overline{PG}_{b}$.

It is assumed that $6 \times 500 MW$ wind farms are available at the network, installed at buses $B_2, B_4, B_5, B_{14}, B_{22}$ and $B_{24}$. Also, a 24 hours operation horizon is considered and the hourly electricity demand and available wind power data are provided in Table \ref{data1}. The nodal distribution of hourly active and reactive power demands is the same with the base data given in \cite{zimmerman2010matpower}. Also, the same profile is assumed for all WFs as given in Table \ref{data1}. Without loss of generality, it is assumed that WFs and P2H electrolyzers are operating with unity power factor in the entire horizon. \textcolor{black}{Besides, it is assumed that demand and generation increment patterns from COP to SLP are through $K^{P/Q}_{b}=1$ for all loads (i.e. with constant power factor) and $K^{G}_{b}=1$ for all generators, except the slack bus generator (i.e. generator $G_2$ at bus $B_{31}$), which is free to increase its active power output in the interval defined in \eqref{lf7}, for providing an hourly balance between total generation and load as well as to compensate the network loss increase.}
The proposed model which is a non-linear programming (NLP) problem, is implemented in the general algebraic modeling system (GAMS) \cite{soroudi2017power} and solved by KNITRO \cite{byrd2006k} solver. $\eta_{b,t}$ in \eqref{p2h2} is assumed to be $13.90 {kg}/{MWh}$ by taking into account the electrolyzer and hydrogen compressor unit energy consumption.

\begin{table}[!t]
	\renewcommand{\arraystretch}{1.3}
	\caption{Hourly power demands and available wind power.}
	\centering
		\label{data1}
	\scalebox{0.65}{
	\begin{tabular}{cccc}
		\hline
	Time	& Active power demand  & Reactive power demand & WF's Available power \\
			&  ($MW$)  &  ($MVAr$) &  ($MW$) \\
		\hline
$t_{1}$	&	5022	&	1114	&	253	\\
$t_{2}$	&	4678	&	1038	&	263	\\
$t_{3}$	&	4472	&	992	&	295	\\
$t_{4}$	&	4018	&	891	&	299	\\
$t_{5}$	&	3770	&	836	&	331	\\
$t_{6}$	&	4087	&	906	&	348	\\
$t_{7}$	&	4541	&	1007	&	367	\\
$t_{8}$	&	5091	&	1129	&	349	\\
$t_{9}$	&	5160	&	1144	&	406	\\
$t_{10}$	&	5917	&	1312	&	455	\\
$t_{11}$	&	6260	&	1388	&	468	\\
$t_{12}$	&	6604	&	1465	&	498	\\
$t_{13}$	&	6329	&	1404	&	490	\\
$t_{14}$	&	5710	&	1266	&	473	\\
$t_{15}$	&	5504	&	1221	&	441	\\
$t_{16}$	&	5641	&	1251	&	404	\\
$t_{17}$	&	5848	&	1297	&	372	\\
$t_{18}$	&	6192	&	1373	&	402	\\
$t_{19}$	&	6467	&	1434	&	409	\\
$t_{20}$	&	6549	&	1453	&	417	\\
$t_{21}$	&	6673	&	1480	&	446	\\
$t_{22}$	&	6880	&	1526	&	500	\\
$t_{23}$	&	6260	&	1388	&	461	\\
$t_{24}$	&	5710	&	1266	&	418	\\
		\hline
	\end{tabular}}
\end{table}

\subsection{Results and discussions}
The impact of three main factors are studied on green hydrogen harvest, namely the optimal allocation of P2H, hourly wind power penetration, and the system LM levels. 

\begin{figure}[th]
  \centering
  \includegraphics[width=0.99\columnwidth,bb=-300 200 900 580,angle=0, clip=true]{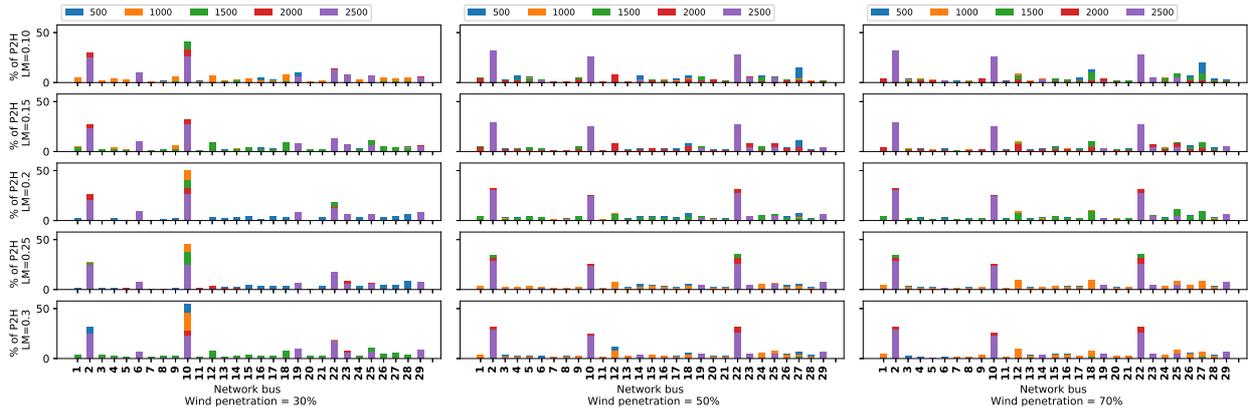} 
  \caption{\textcolor{black}{Optimal P2H allocation for wind penetration ($\alpha$) and LM levels.}}
      \label{fig:ALLOC}
\end{figure}

At first, the problem of optimal allocation of P2H units in the network, including determination of their location and size, is solved by considering all load buses of the studied system as the potential candidates, namely $\Omega_{B_{P2H}}=\left\{ B_{1-29}\right\}$. For this aim, the location and size of P2H demands are obtained for different wind penetration and LM levels. As it is also depicted in Fig. \ref{fig:ALLOC}, by increasing $\alpha$ from $30\%$ to $70\%$, the optimal location of P2H units are determined, and by varying the LM from $10\%$ to $30\%$, the size of P2H units is obtained. 

\begin{figure}[th]
  \centering
  \includegraphics[width=0.8\columnwidth,angle=0, clip=true]{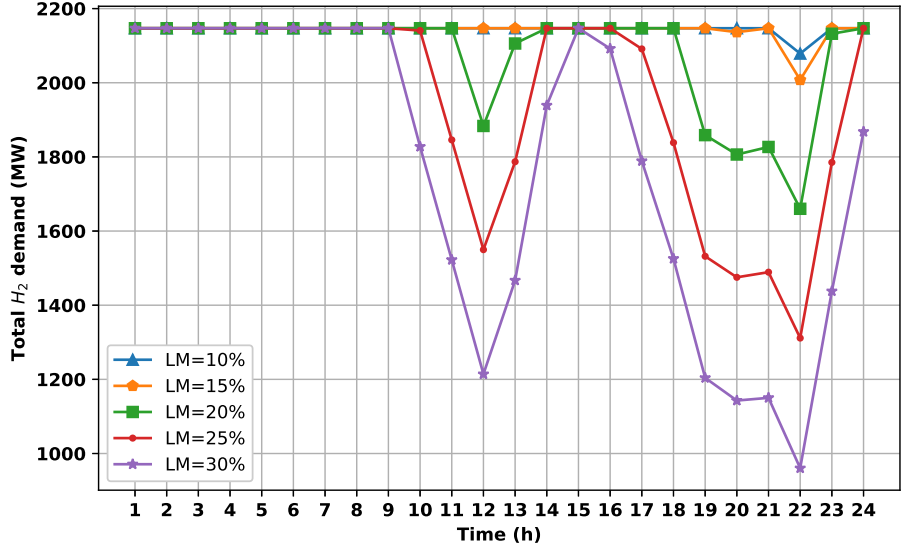} 
      \caption{Hourly total P2H demand for different LM levels ($\alpha=0.50$).}
      \label{fig:THLMhourly}
\end{figure}

\begin{figure}[th]
  \centering
  \includegraphics[width=0.8\columnwidth,angle=0, clip=true]{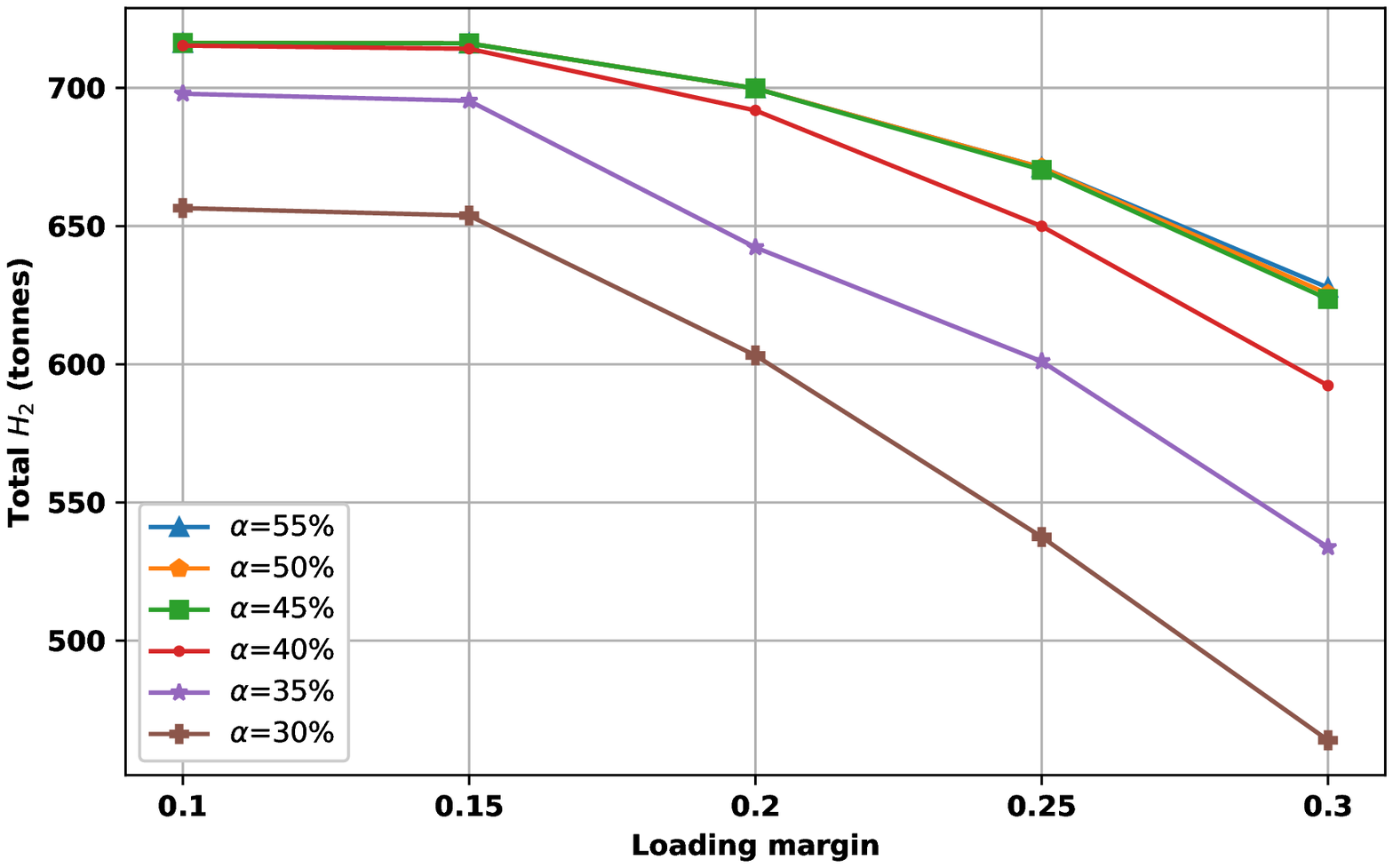} 
      \caption{Total hydrogen vs LM at different wind penetration levels.}\label{fig:THLM}
\end{figure}

It is observed from Fig. \ref{fig:ALLOC} that the optimal buses for P2H units are $B_2$, $B_{10}$, and $B_{22}$ for all LM levels. Also for these locations, the optimal capacity of P2H units will be $805 MW$, $626 MW$, and $716 MW$, respectively. It is also inferred from this figure that for the wind penetration greater than $50\%$, the location and size of P2H units already remain constant. This substantiate the fact that the energy generated by RESs can be converted to green hydrogen only up to a certain level of their penetration and P2H capability will be saturated for higher levels of RESs penetration. From the security point of view, the size of P2H units has an inverse relation with LM, such that smaller P2H capacities can be installed for higher LM values. 

\textcolor{black}{It can be observed from Fig. \ref{fig:ALLOC} that the P2H units are already in the same locations as WFs (at buses $B_2$ and $B_{22}$). It means that for this specific network, these units should be installed near the WFs, but some additional points should be considered for the allocation of P2H units as follows:}
\begin{itemize}
    \item \textcolor{black}{In the event of no available RES power (here, the WFs) in their vicinity (e.g. no wind power in a specific non-windy day), the P2H units should be supplied via the network, may be by the rest of energy producers or via energy import from the external grids.}
    \item \textcolor{black}{The electrolyzes are the new flexible electricity demand that their optimal locations as well as size, may not be the same as the RES generators. By this flexibility, the RES curtailment can be reduced, as well as the issues raised by the injection of large-scale RESs energy to the grid such as congestion, overload in case of contingencies and etc.}
    \item \textcolor{black}{The electrolyzes can be considered as a power flow flexibility option. For example, during the high-wind/low-load condition, their demands help to absorb more RESs power to the grid.}
    \item \textcolor{black}{Transmission of electric power is much faster and easier than hydrogen transportation. For large-scale applications, it is more convenient to install the electrolyzes near the main hydrogen demands rather that WFs, as the electric power is readily in hand via the electricity transmission network.}
\end{itemize}

\textcolor{black}{Also, the impact of LM on the hourly attainable green hydrogen is presented in Fig. \ref{fig:THLMhourly} for a give value of wind penetration (i.e. $\alpha=0.50$). It is evident from this figure that, for the aforementioned allocation of P2H units, the total P2H demand of the system is decreasing for higher levels of LM. It is observed from this figure that in the off-peak periods such as $t_1-t_9$, there is no security imposed limitation on the energy consumption of P2H units, and all three units are operated at their corresponding nominal ratings. But at peak demand periods, such as the intervals $t_{10}-t_{14}$ or $t_{16}-t_{24}$, security limit constraints restricted the energy conversion to green hydrogen.}
Finally, the total hydrogen extracted for different levels of LM and wind penetration is shown in Fig. \ref{fig:THLM}. 
It is observed from this figure that for a given wind penetration level the voltage security constraints are not binding up to a certain level of LM (here $15\%$), and the hydrogen extraction is limited by the remaining constraints such as penetrable wind and the energy flow equations of the network. On the other hand, for LMs higher than $15\%$, the voltage security constraints are activated such that the attainable green hydrogen decreases. Besides, for a given value of LM, by decreasing the wind penetration level, the amount of exploitable green hydrogen is also decreased. Since the P2H units can by supplied through the network and remaining energy sources (such as thermal generation units), this reduction in P2H demand with respect to decrements of wind penetration, validates the proper location of P2H units to convert the wind power to green hydrogen.

\section{Conclusion}
\label{sec:Conclusion}
This paper explores the most influential technical factors affecting the large-scale green hydrogen harvest. It is demonstrated that the optimal location of P2H units, loadability constraints as well as the RESs penetration can pose considerable impacts on the exploitable green hydrogen. The model is implemented on the IEEE 39-bus system and the concluding remarks can be summarized as follows:
\begin{itemize}
   \item The optimal location of P2Hs mainly depends on penetration level of RESs, whereas the size of P2Hs plants has an inverse relation with the LM levels.
   \item For a given size of P2H units, the harness-able green hydrogen decreases in higher levels of LM.
   \item At higher wind penetration levels more green hydrogen is available, for a give level of LM. Beyond a certain threshold, LM constraints restrict the exploitable green hydrogen.
\end{itemize}

Suggestions for future work:
\begin{itemize}
    \item The impact of hydrogen extraction on gas networks should be investigated. A portion of the electricity demand will be supplied via thermal power plants. These generators are supplied by gas networks. The technical and economic set-points of gas networks will be affected. \item A more detailed security constrained AC-OPF can better characterize the impact of hydrogen extraction in power systems. 
    \item The risk and uncertainty of renewable energy resources should be taken into account to avoid financial and technical risks \cite{zhu2020optimal}. \end{itemize}

\vspace{5mm}
\section*{Acknowledgements}
The work done by Alireza Soroudi is supported by a research grant from Science Foundation Ireland (SFI) under the SFI Strategic Partnership Programme Grant No. SFI/15/SPP/E3125. The opinions, findings and conclusions or recommendations expressed in this material are those of the author(s) and do not necessarily reflect the views of the Science Foundation Ireland. 
\bibliography{MP-ref}

\begin{thebibliography}{10}
\expandafter\ifx\csname url\endcsname\relax
  \def\url#1{\texttt{#1}}\fi
\expandafter\ifx\csname urlprefix\endcsname\relax\def\urlprefix{URL }\fi
\expandafter\ifx\csname href\endcsname\relax
  \def\href#1#2{#2} \def\path#1{#1}\fi

\bibitem{staffell2019role}
I.~Staffell, D.~Scamman, A.~V. Abad, P.~Balcombe, P.~E. Dodds, P.~Ekins,
  N.~Shah, K.~R. Ward, The role of hydrogen and fuel cells in the global energy
  system, Energy \& Environmental Science 12~(2) (2019) 463--491.

\bibitem{wang2020well}
Q.~Wang, M.~Xue, B.-L. Lin, Z.~Lei, Z.~Zhang, Well-to-wheel analysis of energy
  consumption, greenhouse gas and air pollutants emissions of hydrogen fuel
  cell vehicle in china, Journal of Cleaner Production (2020) 123061.

\bibitem{IRENA1}
IRENA, Hydrogen from renewable power: Technology outlook for the energy
  transition, Tech. rep., International Renewable Energy Agency. URL:
  https://www.irena.org/publications/2018/Sep/Hydrogen-from-renewable-power
  (2018).

\bibitem{jiao2020green}
J.~Jiao, C.~Chen, Y.~Bai, Is green technology vertical spillovers more
  significant in mitigating carbon intensity? evidence from chinese industries,
  Journal of Cleaner Production 257 (2020) 120354.

\bibitem{FASIHI2020118466}
M.~Fasihi, C.~Breyer, Baseload electricity and hydrogen supply based on hybrid
  pv-wind power plants, Journal of Cleaner Production 243 (2020) 118466.

\bibitem{eichman2016economic}
J.~Eichman, A.~Townsend, M.~Melaina, Economic assessment of hydrogen
  technologies participating in california electricity markets, Tech. rep.,
  National Renewable Energy Lab.(NREL), Golden, CO (United States) (2016).

\bibitem{hou2017optimizing}
P.~Hou, P.~Enevoldsen, J.~Eichman, W.~Hu, M.~Z. Jacobson, Z.~Chen, Optimizing
  investments in coupled offshore wind-electrolytic hydrogen storage systems in
  denmark, Journal of Power Sources 359 (2017) 186--197.

\bibitem{kopp2017energiepark}
M.~Kopp, D.~Coleman, C.~Stiller, K.~Scheffer, J.~Aichinger, B.~Scheppat,
  Energiepark mainz: Technical and economic analysis of the worldwide largest
  power-to-gas plant with pem electrolysis, International Journal of Hydrogen
  Energy 42~(19) (2017) 13311--13320.

\bibitem{mansour2020robust}
A.~Mansour-Saatloo, M.~Agabalaye-Rahvar, M.~A. Mirzaei, B.~Mohammadi-Ivatloo,
  K.~Zare, et~al., Robust scheduling of hydrogen based smart micro energy hub
  with integrated demand response, Journal of Cleaner Production (2020) 122041.

\bibitem{MURRAY2020117792}
P.~Murray, J.~Carmeliet, K.~Orehounig, Multi-objective optimisation of
  power-to-mobility in decentralised multi-energy systems, Energy 205 (2020)
  117792.

\bibitem{IRENA2}
IRENA, Innovation landscape brief: Renewable power-to-hydrogen, Tech. rep.,
  International Renewable Energy Agency. URL:
  https://www.irena.org/-/media/Files/IRENA/Agency/Publication/2019/Feb/IRENA-Innovation-Landscape-2019-report.pdf
  (2019).

\bibitem{6294475}
A.~{Rabiee}, M.~{Parniani}, Voltage security constrained multi-period optimal
  reactive power flow using benders and optimality condition decompositions,
  IEEE Transactions on Power Systems 28~(2) (2013) 696--708.

\bibitem{8606272}
S.~M. {Mohseni-Bonab}, I.~{Kamwa}, A.~{Moeini}, A.~{Rabiee}, Voltage security
  constrained stochastic programming model for day-ahead bess schedule in
  co-optimization of t d systems, IEEE Transactions on Sustainable Energy
  11~(1) (2020) 391--404.

\bibitem{zimmerman2010matpower}
R.~D. Zimmerman, C.~E. Murillo-S{\'a}nchez, R.~J. Thomas, Matpower:
  Steady-state operations, planning, and analysis tools for power systems
  research and education, IEEE Transactions on power systems 26~(1) (2010)
  12--19.

\bibitem{soroudi2017power}
A.~Soroudi, Power system optimization modeling in GAMS, Springer, 2017.

\bibitem{byrd2006k}
R.~H. Byrd, J.~Nocedal, R.~A. Waltz, Knitro: An integrated package for
  nonlinear optimization, in: Large-scale nonlinear optimization, Springer,
  2006, pp. 35--59.

\bibitem{zhu2020optimal}
Y.~Zhu, Q.~Tong, X.~Yan, Y.~Liu, J.~Zhang, Y.~Li, G.~Huang, Optimal design of
  multi-energy complementary power generation system considering fossil energy
  scarcity coefficient under uncertainty, Journal of Cleaner Production (2020)
  122732.

\end{thebibliography}

\end{document}